\documentclass[doublecol,]{epl2}

\begin{document}
\title{Fluidisation and plastic activity in a model soft-glassy material flowing
in micro-channels with rough walls}

\shorttitle{Fluidisation and plastic activity in rough micro-channels}

\author{A. SCAGLIARINI\inst{1,2}, M. LULLI\inst{2}, M. SBRAGAGLIA\inst{2}, AND M. BERNASCHI\inst{3}}
\shortauthor{A. Scagliarini \etal}

\institute{
  \inst{1} Helmholtz-Institute Erlangen-N\"{u}rnberg for Renewable Energy Production \\
               Forschungszentrum J\"{u}lich GmbH (IEK-11) F\"{u}rther Strabe 248, 91429 N\"{u}rnberg, Germany \\
  \inst{2} Department of Physics and INFN, University of Tor Vergata, Via della Ricerca Scientifica 1, 00133 Rome, Italy \\
  \inst{3} Istituto per le Applicazioni del Calcolo CNR, Via dei Taurini 9, 00185 Roma, Italy.}


\abstract{By means of mesoscopic numerical simulations of a model soft-glassy material, we investigate the role of boundary roughness on the flow behaviour of the material, probing the bulk/wall and global/local rheologies. We show that the roughness reduces the wall slip induced by wettability properties and acts as a source of fluidisation for the material. A direct inspection of the plastic events suggests that their rate of occurrence grows with the fluidity field, reconciling our simulations with kinetic elasto-plastic descriptions of jammed materials. Notwithstanding, we observe qualitative and quantitative differences in the scaling, depending on the distance from the rough wall and on the imposed shear. The impact of roughness on the orientational statistics is also studied.}

\maketitle

\section{Introduction}
The general term {\it Soft-Glassy Materials} (SGM) embraces a number of complex materials of great technological and biological relevance, whose rheology lies in between solid-like and liquid-like behaviors \cite{Larson}. Dense emulsions, foams and gels are instances of such systems. Due to their importance in a host of natural and industrial processes, and to the challenge represented by their modelling for non-equilibrium statistical mechanics, SGM have been the subject of many recent experimental \cite{Mason,Experimental}, theoretical \cite{Princen,Theoretical,Falk,KEP} and numerical works \cite{Numerical}. It is widely acknowledged that such materials flow as the result of a succession of plastic rearrangements, occurring when a local configuration of constituting micro-elements (i.e. droplets for emulsions, bubbles for a foam, etc) cannot sustain the accumulated stress and relaxes it in the form of long-ranged elastic waves, which induce non-locality in the rheological properties of the system. A number of theoretical frameworks have been developed recently to take into account these non-local effects \cite{Falk,KEP,Bouzid,KamrinKoval}. One of them, the Kinetic Elasto-Plastic (KEP) model \cite{KEP}, captures the essential phenomenology in a mean-field spirit through a diffusion-relaxation equation for the {\it fluidity} field $f=\dot{\gamma}/\sigma$  (the ratio of the {\it local} shear rate $\dot{\gamma}$ and shear stress $\sigma$):
\begin{equation} \label{eq:fluidity}
\xi^2 \Delta f = f - f_{\mbox{\tiny{b}}},
\end{equation}
where $\xi$ is the so-called {\it cooperativity length}, measuring the spatial extension of the non-local correlations, and $f_{\mbox{\tiny{b}}}=f_{\mbox{\tiny{b}}}(\sigma)$ is the {\it bulk fluidity}, which is a function of the shear stress only and equals the fluidity in absence of spatial heterogeneities. The other fundamental result of KEP, along with (\ref{eq:fluidity}), is the expected proportionality between $f$ and the rate of occurrence of plastic events (whose prototypical $\mbox{2d}$ instance is the so-called $T1$, i.e. the neighbour-swapping of four adjacent droplets/bubbles), $R_{T1}$:
\begin{equation} \label{eq:fluidity-2}
R_{T1} = \mathcal{A} f,
\end{equation}
with $\mathcal{A}$ a constant, proportional to the elastic modulus $G_0$ in the KEP formulation. Equation (\ref{eq:fluidity}) can be solved analytically in some simple cases, provided that a proper boundary condition is supplied. In a Couette flow, for instance, in which the stress $\sigma$ is constant, a straightforward calculation for a Dirichlet-type boundary condition $f(\pm H/2) = f_{\mbox{\tiny{w}}}$ yields
\begin{equation}
f(z)=f_{\mbox{\tiny{b}}} + (f_{\mbox{\tiny{w}}} - f_{\mbox{\tiny{b}}})\frac{\cosh(z/\xi)}{\cosh(H/(2\xi))},
\end{equation}
where $-H/2 \le z \le H/2$ is the distance from the bottom ($z=-H/2$) wall and $H$ the wall-to-wall distance. The corresponding velocity profile $v(z) = \sigma \int f(\zeta) d\zeta$ has been found to be consistent with numerical and experimental data \cite{SM2014,KamrinKoval}. However, a Von Neumann boundary condition of the type $\frac{d  f}{dz}(\pm H/2) = 0$ (i.e. the wall does not act as a ``source'' of fluidity) would give $f=f_{\mbox{\tiny{b}}}$ everywhere, in clear contrast with all known facts. It must be noticed that, though it plays a crucial role, the fluidity at the wall $f_{\mbox{\tiny{w}}}$ has no first-principles ground and, as a matter of fact, enters the picture as a completely phenomenological free parameter of the model. Therefore, a challenging question, which certainly deserves thourough investigations, concerns the possibility of engineering a surface in such a way to control the fluidity, and hence the plastic activity, in its proximity. So far, only very few studies addressed the problem. Mansard {\it et al.} \cite{Mansard14} performed experiments with dense emulsions flowing in micro-channels with rough walls patterned with equally spaced posts of variable height, finding that slippage and surface fluidization depend non-monotonously on it. By means of lattice Boltzmann (LB henceforth) numerical simulations of Poiseuille flows of model soft-glassy materials \cite{ourJSP}, we have recently shown that there is a non-trivial dependence also on other geometrical parameters of the roughness (namely, the inter-post distance); additionally, the presence of the roughness may cause a breakdown of the proportionality relation predicted by KEP between fluidity and rate of occurrence of plastic events. To address the latter problem in a more systematic way, in this paper we study $\mbox{2d}$ Couette flows. Starting from a smooth boundary, and exploring different realizations of the roughness, we are going to assess how the fluidity field changes close to the boundaries, hence providing the boundary condition for the bulk-flow. Roughness effects are materialized through {\it enhanced friction} and {\it surface fluidization} induced by enhanced plastic activity. Numerical results are rationalized in the framework of KEP theory \cite{KEP} for elastoplastic materials.


\section{Model}

\begin{figure}
\begin{center}
\advance\leftskip-0.55cm
\includegraphics[scale=0.4]{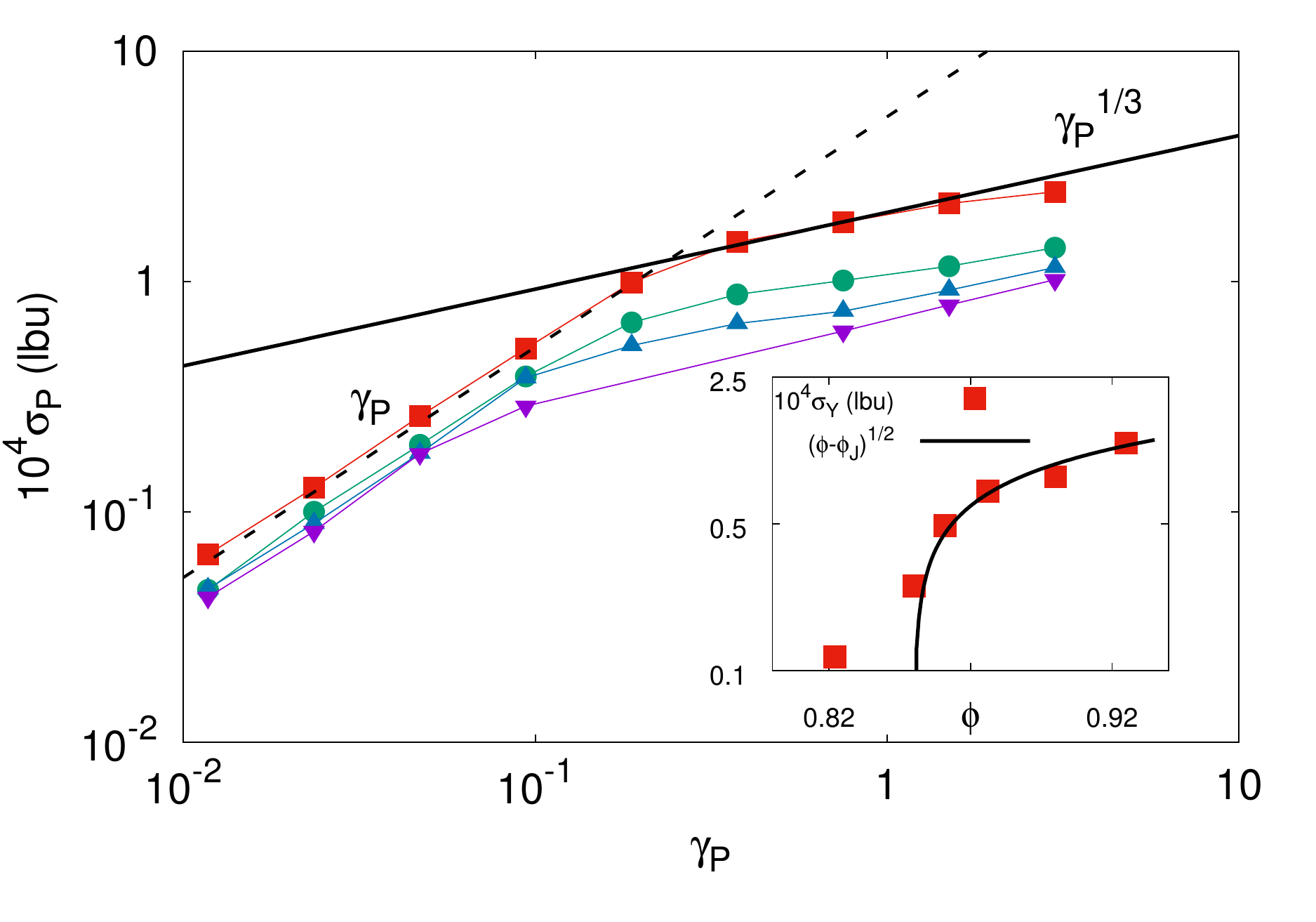}
\includegraphics[scale=0.62]{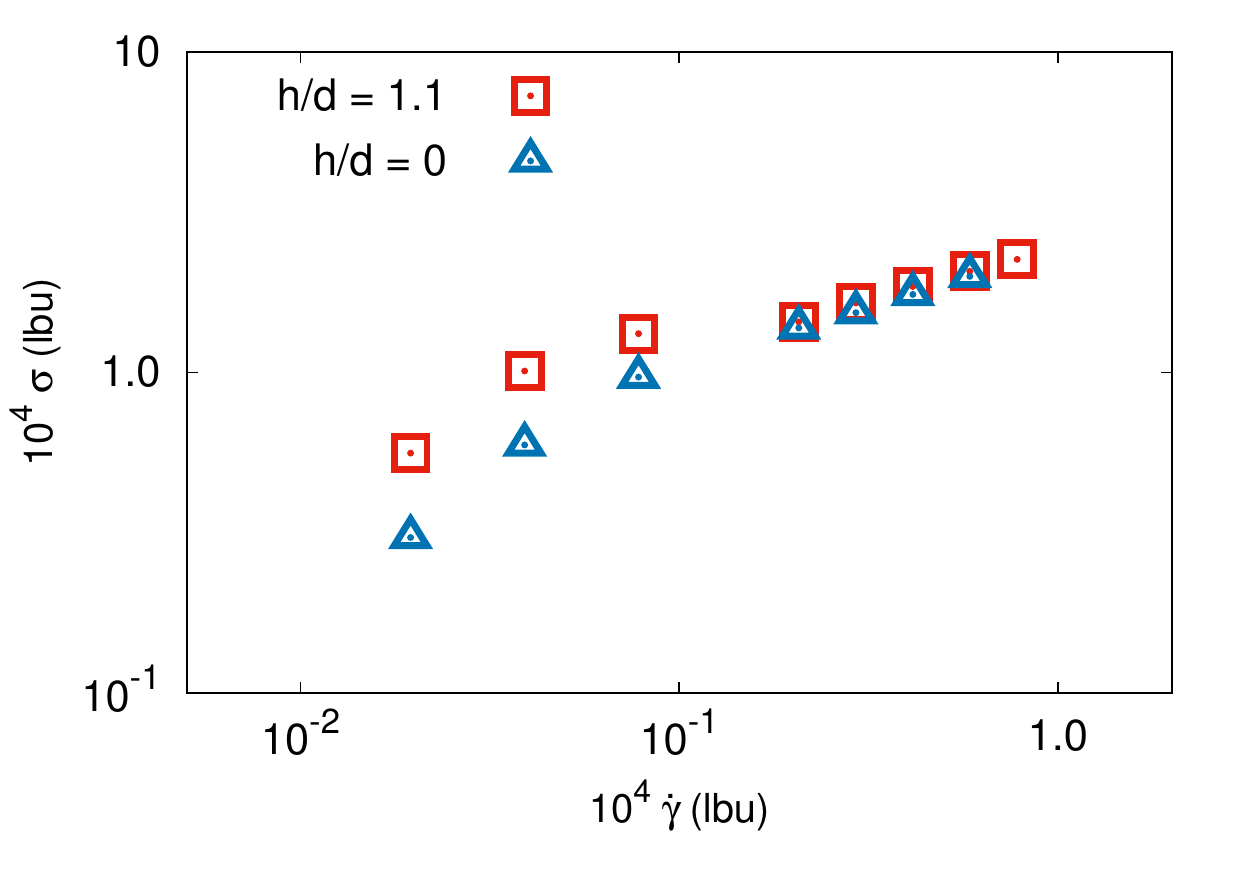}
\caption{ Top panel: flow curves (peak stress $\sigma_P$ {\it vs.} peak strain $\gamma_P$) from numerical simulations of oscillating Couette flows (with smooth walls) for various volume fractions. The dashed line indicates the linear behaviour below the yield stress, whereas the solid one is the rheo-thinning law $\gamma_P^{1/3}$. Inset: Yield stress $\sigma_Y$ as a function of the volume fraction of the dispersed phase. Bottom panel: flow curve obtained in a steady Couette flow with smooth ($h/d=0$) and rough ($h/d= 1.1$) bottom wall. \label{fig:Sigma_vs_Pxy}}
\end{center}
\end{figure}

The numerical model used here was described in several previous works \cite{LBGLASSY}. It is a mesoscopic LB for non ideal binary fluids allowing the description of a collection of droplets dispersed in a continuous phase. The coalescence of these droplets is inhibited by the (positive) disjoining pressure which develops in the thin films separating two approaching interfaces. The model gives direct access to the hydrodynamical variables (expressed in this paper in lattice Boltzmann units, lbu), i.e. density and velocity fields, as well as the local (in time and space) stress tensor in the system. Thus, it is extremely useful to characterize the relationship between the droplets dynamics, their plastic rearrangements, and the stress fluctuations. Once the droplets are stabilized, different packing fractions and polydispersities can be obtained, and the model develops a yield stress, as evident from the curves displayed in the top panel of Figure \ref{fig:Sigma_vs_Pxy}. Specifically, we performed rheometric tests, submitting samples of our soft-glassy material with different composition ratios to an oscillatory strain; this was achieved setting the velocity of the top wall to $U_{\mbox{\tiny{w}}}(t) = U_{\mbox{\tiny{w}}}^{({\mbox{\tiny{max}}})} \sin(\omega t)$, with the pulsation $\omega$ small enough to probe also the viscous (liquid-like) response at spanning the applied strain by changing $U_{\mbox{\tiny{w}}}^{({\mbox{\tiny{max}}})}$. In such rheometric tests no geometric roughness is present (both walls are smooth); a neutral wetting condition for both the continuous (c) and dispersed (d) phases (contact angle $\theta_{\mbox{\tiny{d}}}=\theta_{\mbox{\tiny{c}}} = 90^{\circ}$) is imposed on both walls. Notice that, without roughness, we cannot set a complete wetting condition $\theta_{\mbox{\tiny{c}}}=0^{\circ}$ for the continuous phase (as we will do later on), since this would give rise to slip effects that may spoil rheological measurements \cite{Bonn2010,Bonn2015}. Our numerical experiment echoes the (real-world) experiments by Mason {\it et al.} \cite{Mason}: in Figure \ref{fig:Sigma_vs_Pxy} (top panel) we plot the measured peak shear stress $\sigma_P$ developed inside the material as a function of the maximum imposed strain $\gamma_P$, for various values of the volume fraction of the dispersed phase $\phi$. We observe that the system displays distinctive features of soft-glassy rheology, namely it develops a yield point, below which it behaves elastically ($\sigma_P \propto \gamma_P$) and above it shows a sublinear behaviour in the relation between $\sigma_P$ and $\gamma_P$. The value of the yield stress $\sigma_Y$ decreases with $\phi$ and eventually must be expected to vanish at the {\it jamming} value $\phi_J$, since it is the close packing which confers rigidity to the system. As a matter of fact, this is what we see in the inset of Figure \ref{fig:Sigma_vs_Pxy}, where $\sigma_Y$ is plotted against $\phi$: the yield stress goes to zero at $\phi_J$ as $(\phi - \phi_J)^{1/2}$ (solid line). The $\phi^{1/2}$ dependence of the yield stress on the volume fraction was predicted theoretically, on the basis of purely geometrical arguments, by Princen for dry foams (and highly concentrated emulsions) \cite{Princen}. 



\section{Numerical simulations with rough walls}

\begin{figure}
\begin{center}
\advance\leftskip-0.55cm
\includegraphics[scale=0.3]{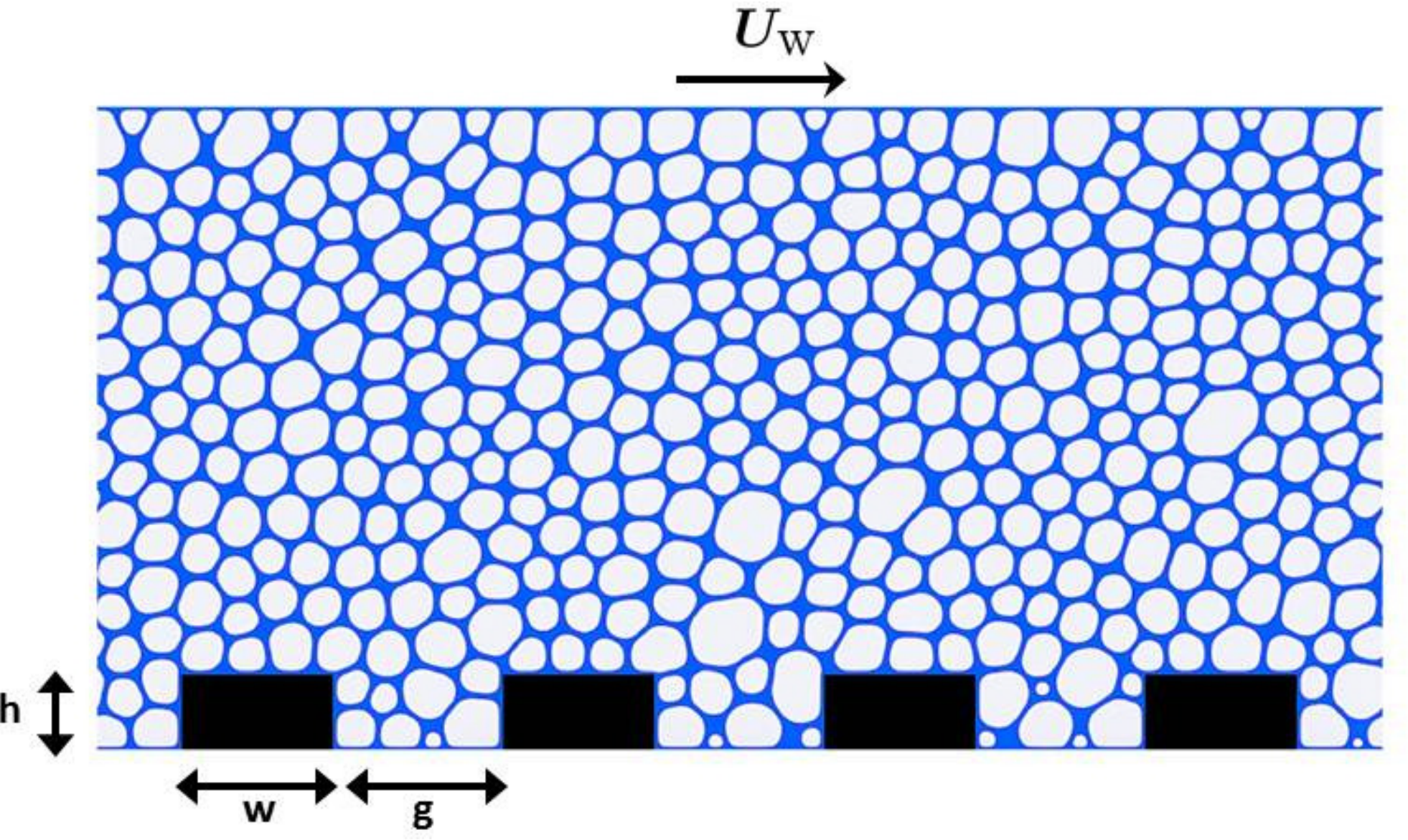}
\includegraphics[scale=0.4]{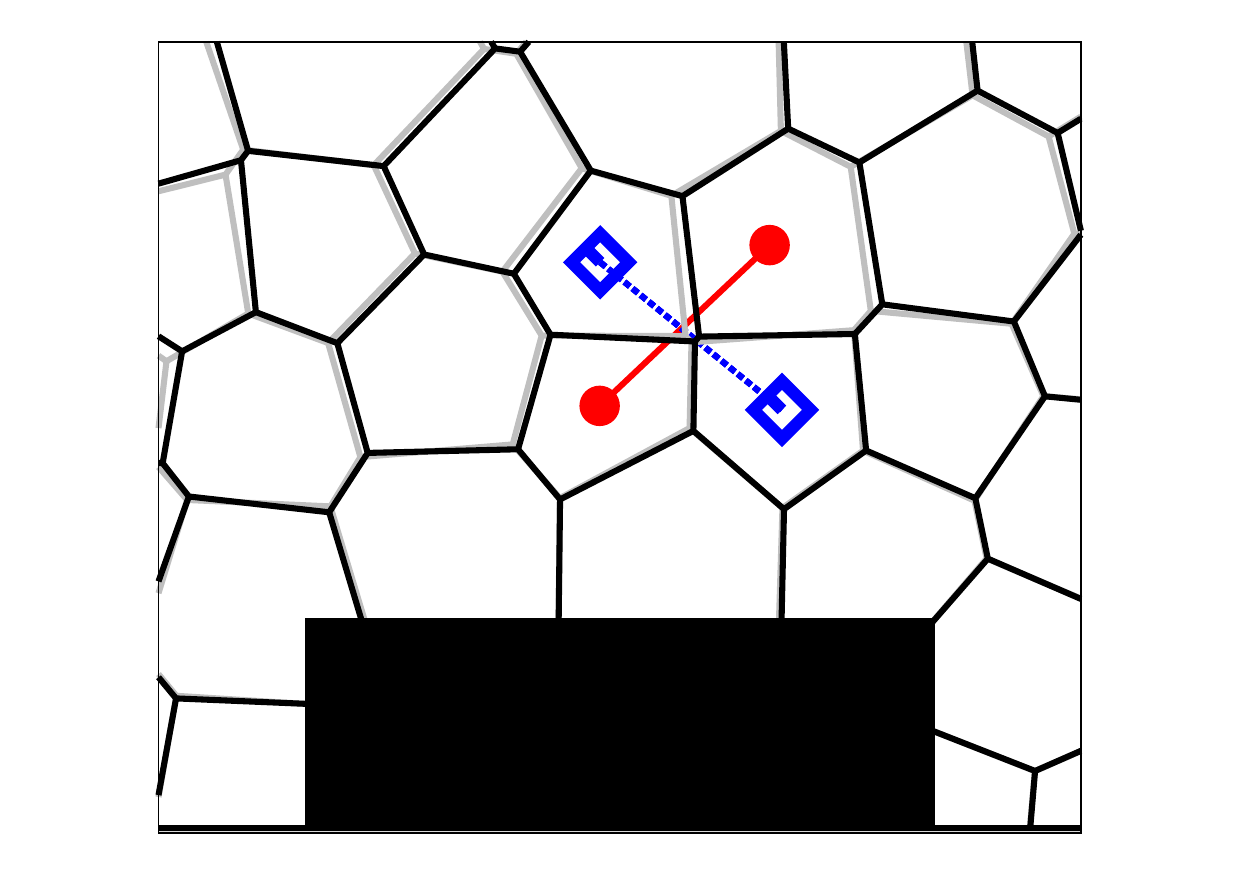}
\caption{Top panel: a snapshot of the density field (light color represents the dispersed phase) from LB simulations \cite{LBGLASSY} in confined micro-channels with structured roughness on the bottom wall: $g$, $w$ and $h$ will refer to the inter-post distance and the post width and height, respectively. Bottom panel: Sketch of a T1 plastic event. To analyze plastic events, we perform a Voronoi tessellation from the centers of mass of the droplets. Following the Voronoi tessellation in time, we are able to identify T1 events and associated ``disappearing'' (red solid line) and ``appearing'' (blue dashed line) links. In grey (black) we indicate the Voronoi cells soon before (after) a T1 event.
}\label{sketch}
\end{center}
\end{figure}

Our numerical setup is sketched in Figure \ref{sketch}, where we also show a snapshot of the density map taken from a simulation. The soft-glassy material is confined between a top wall moving with velocity $U_{\mbox{\tiny{w}}}$  and a fixed bottom wall patterned with equally spaced posts of width $w$, height $h$ and inter-post distance $g$; all these lengths are given in units of the mean droplet diameter $d$. All the numerical simulations have been carried out in domains resolved with $L_x \times H = 1024 \times 512$ lbu. With a single droplet diameter resolved with roughly $30$-$35$ lbu, we are able to accommodate roughly $20$ droplets diameters in the wall-to-wall gap $H$ (see also figure \ref{sketch}). Since long execution times may be expected in order to accumulate a reliable statistics on such domains, we resorted to Graphic Processing Units (GPUs). We mostly used Nvidia Kepler ``K80'' cards that feature two GPUs each one having a total of 2496 cores and a memory bandwidth of 240 GB/sec. On such platform it is possible to efficiently simulate systems of linear sizes of the order of a few thousands of lattice points. The code we used extends and enhances our original CUDA implementation of the LB glassy model described in \cite{PREGPU}. A typical run of $80 \times 10^{6}$ time steps, which takes 10 days on a single GPU, produces about $10^{4}$ statistically independent configurations, with a total of about $5 \times 10^4$ plastic rearrangements counted over the whole domain.


\begin{figure}
\begin{center}
\advance\leftskip-0.55cm
\includegraphics[scale=0.43]{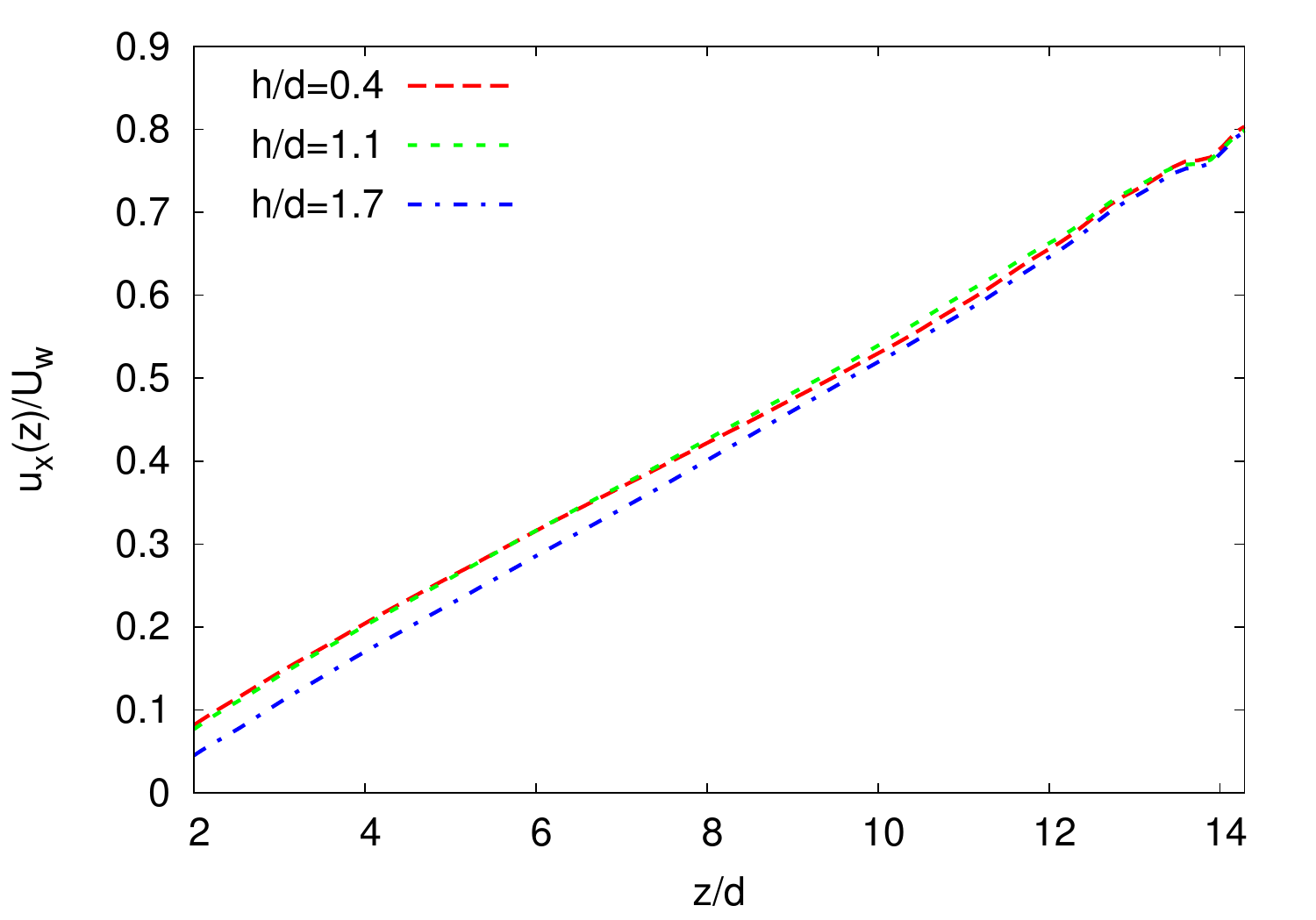}
\includegraphics[scale=0.43]{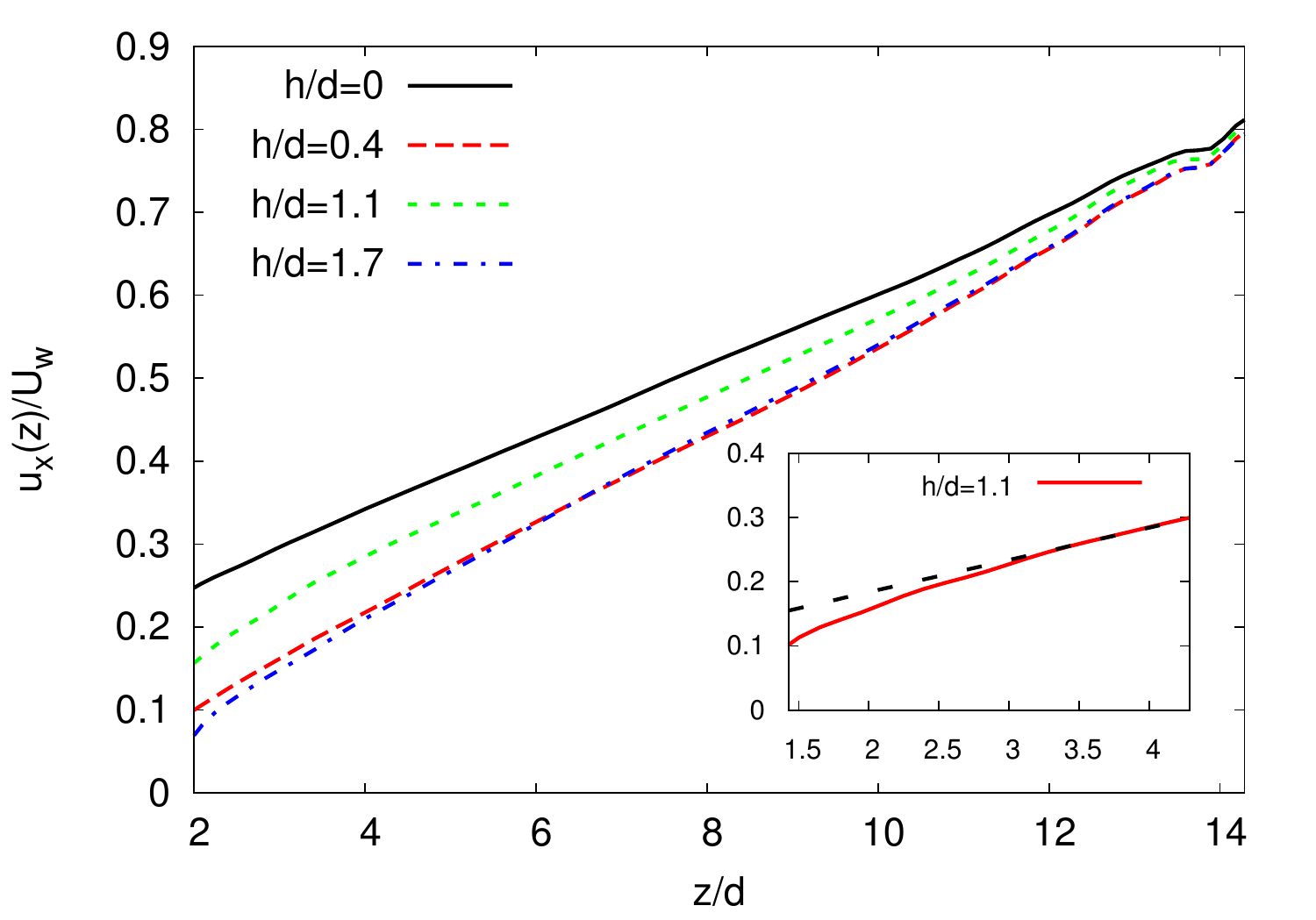}
\caption{Stream-wise velocity profiles $u_x(z)$ for $U_{\mbox{\tiny{w}}}=0.015$ lbu and different realizations of the roughness parameters: $g/d=6.7$, $w/d=0.6$, $h/d=0.4,1.1,1.7$ (top panel) and $g/d=4.0$, $w/d=3.5$, $h/d=0,0.4,1.1,1.7$ (bottom panel). Inset: zoom of the velocity profile close to the roughness edge for $h/d=1.1$, showing deviations from the linear (dashed line) behaviour. \label{velprofs-L256-W120}}
\end{center}
\end{figure}

\begin{figure}
\begin{center}
\advance\leftskip-0.55cm
\includegraphics[scale=0.45]{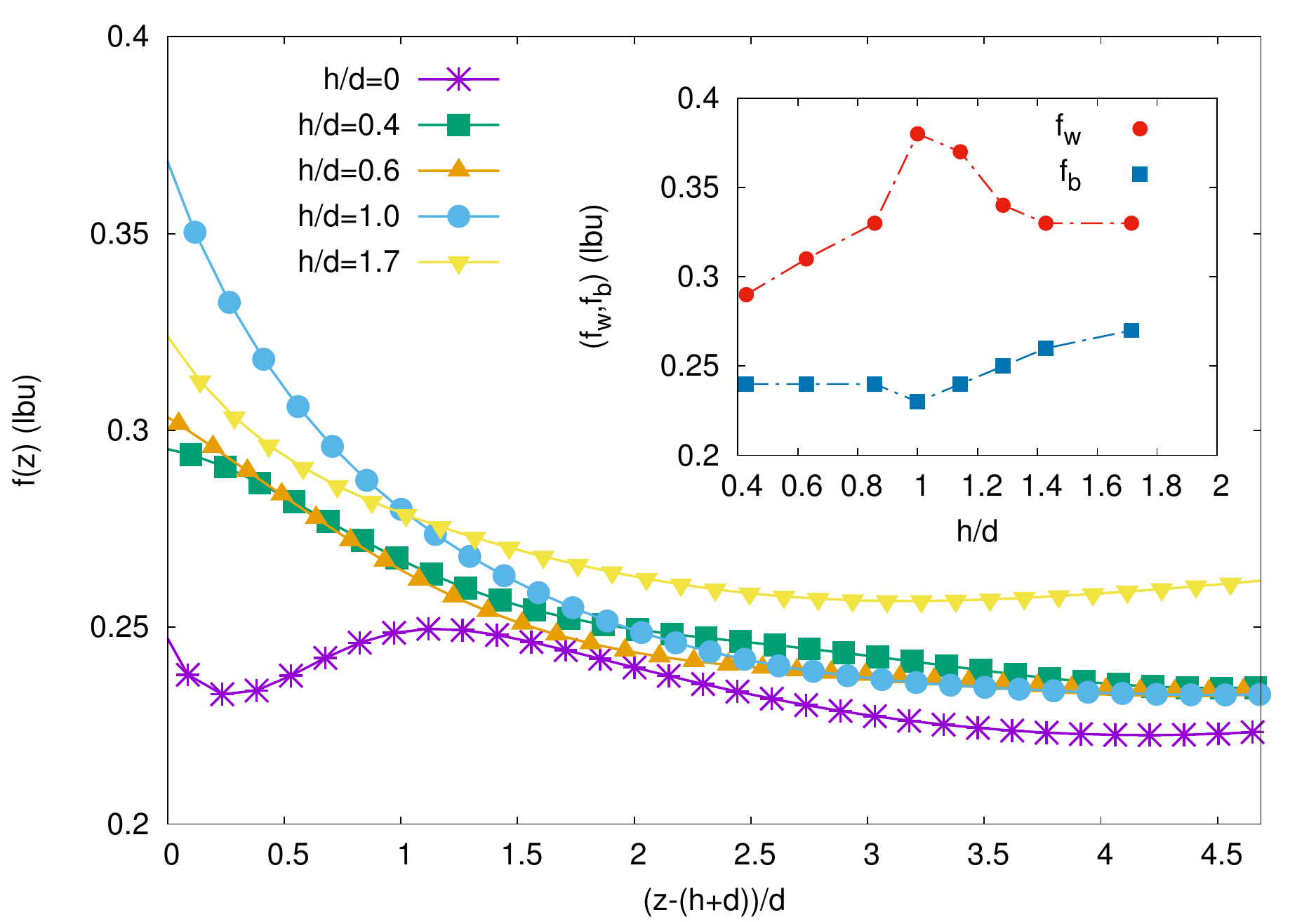}
\caption{Fluidity profiles for $g/d=4.0$, $w/d=3.5$ and various $h/d$. Each profile is shifted in the $z$-axis in such a way that the origin corresponds to a location at a distance $d$ from the edge of the posts. Inset: wall ($f_{\mbox{\tiny{w}}}$) and bulk fluidities ($f_{\mbox{\tiny{b}}}$), for a fixed top wall velocity, as a function of the post height.\label{f-l256-w120}}
\end{center}
\end{figure}

\begin{figure}
\begin{center}
\advance\leftskip-0.55cm
\includegraphics[scale=0.5]{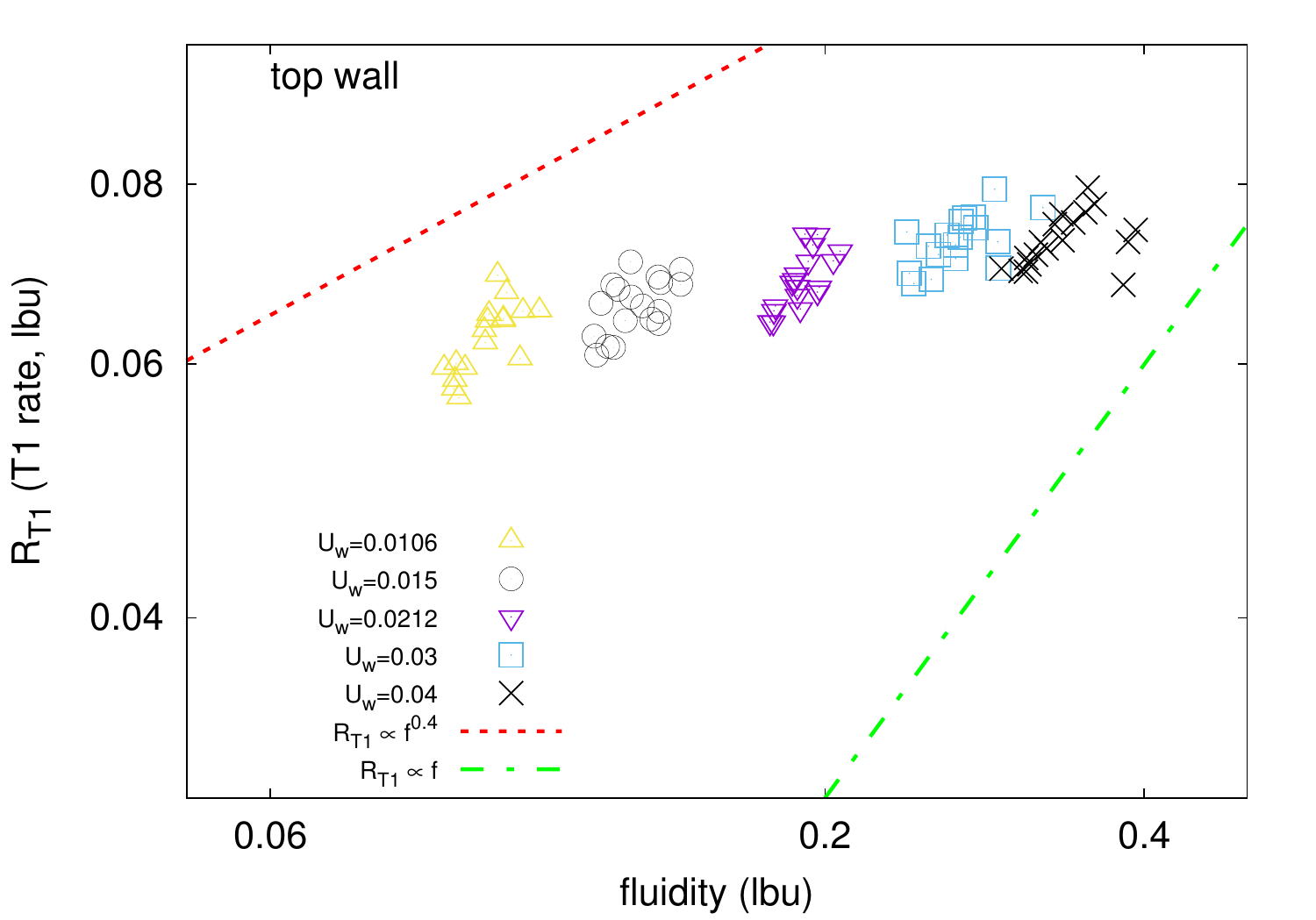}
\includegraphics[scale=0.5]{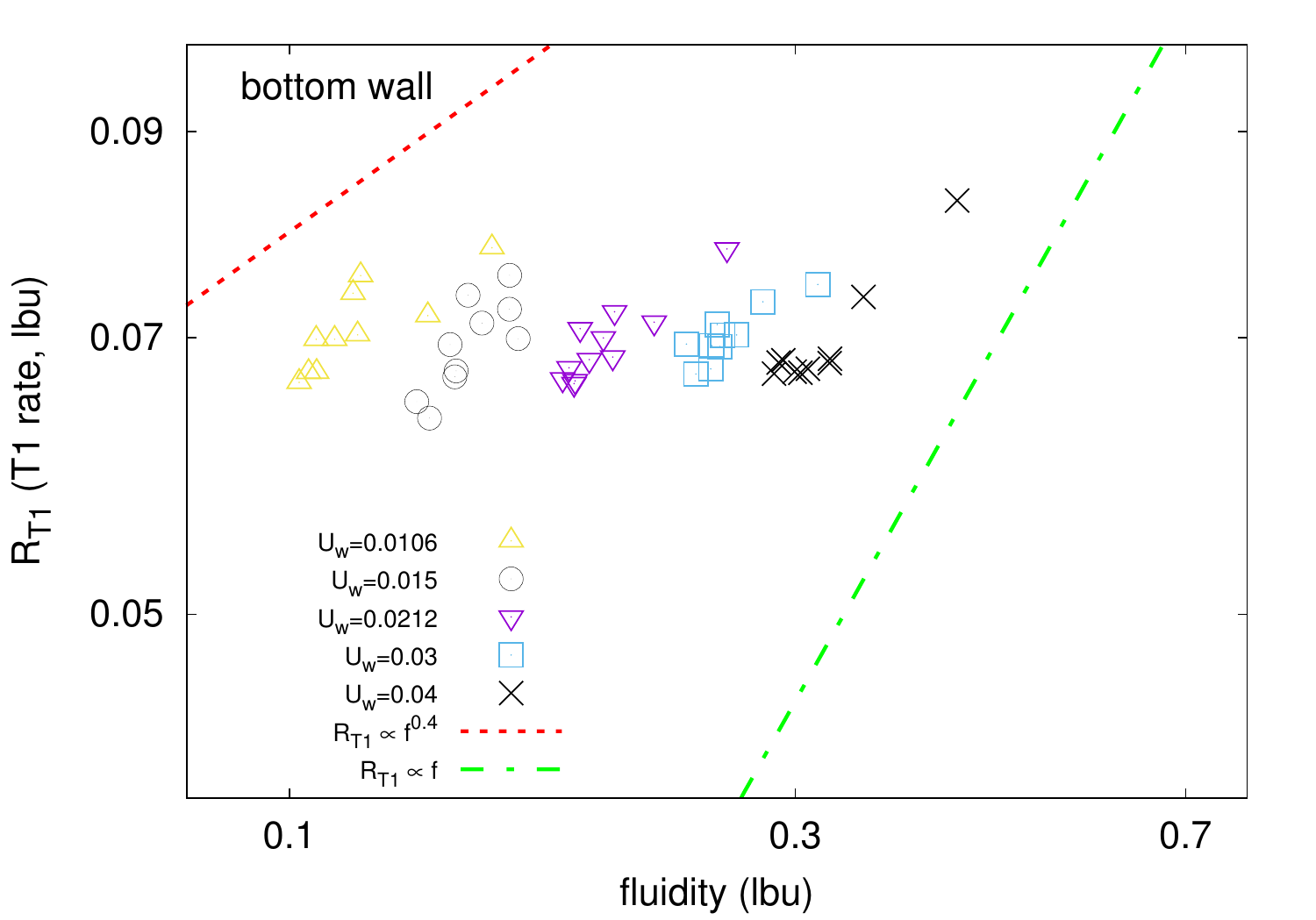}
\caption{(Colour online). Scatter plot of T1 rate {\it vs.} fluidity computed within two slabs of width $\sim 3\,d$, close to the top wall (top panel) and bottom wall (bottom panel), respectively. Each family of points corresponds to a certain value of the top wall velocity $U_{\mbox{\tiny{w}}}$ (given in lbu), with roughness parameters $g/d=4.0$, $w/d=3.5$, $h/d=0,0.4,1.1,1.7$. Two lines, corresponding to the scalings $R_{T1} \propto f$ (dashed-dotted line) and $R_{T1} \propto f^{0.4}$ (dotted line), are drawn to guide the eye. \label{scatter}}
\end{center}
\end{figure}

\begin{figure}
\begin{center}
\advance\leftskip-0.55cm
\includegraphics[scale=0.5]{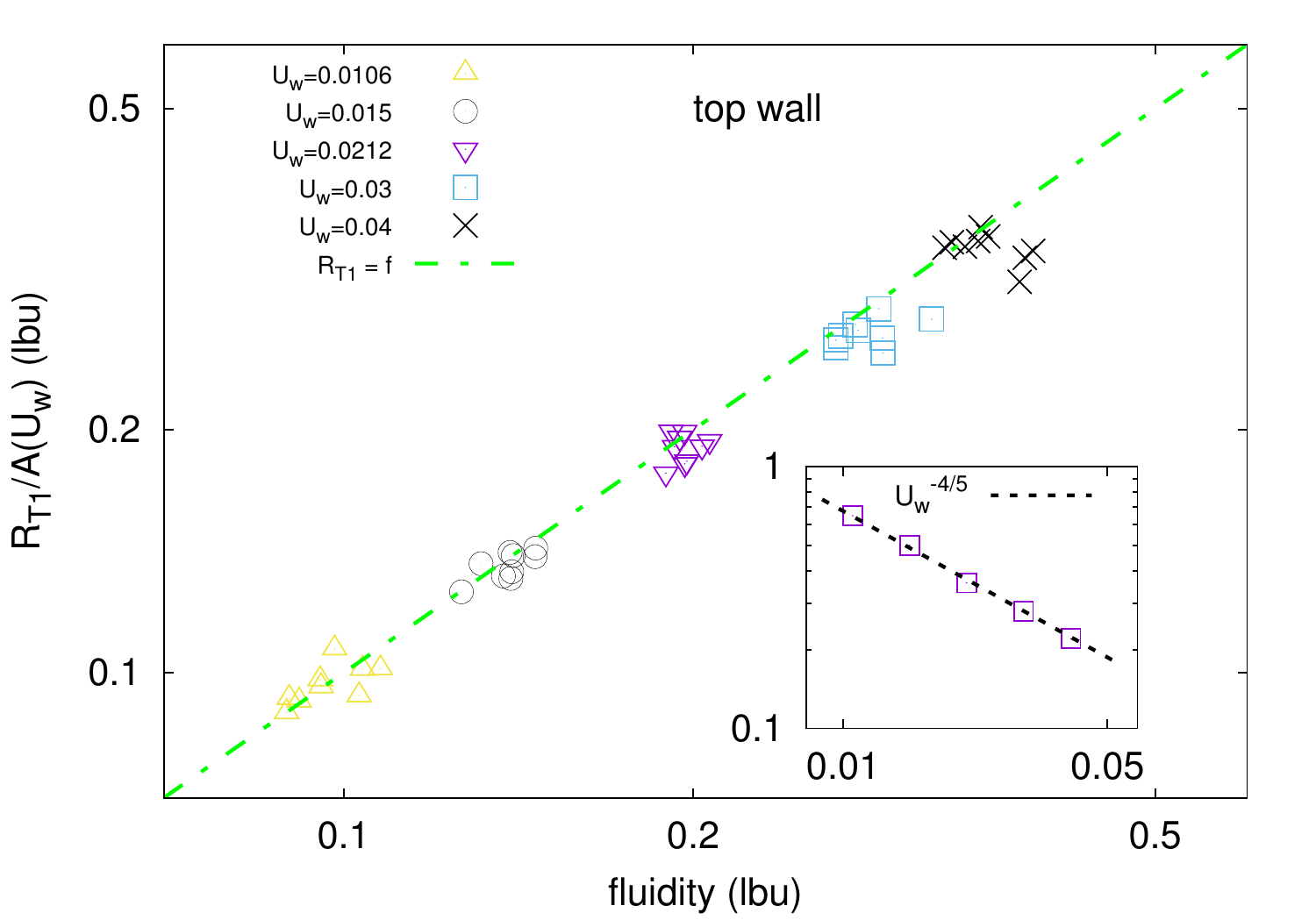}
\includegraphics[scale=0.5]{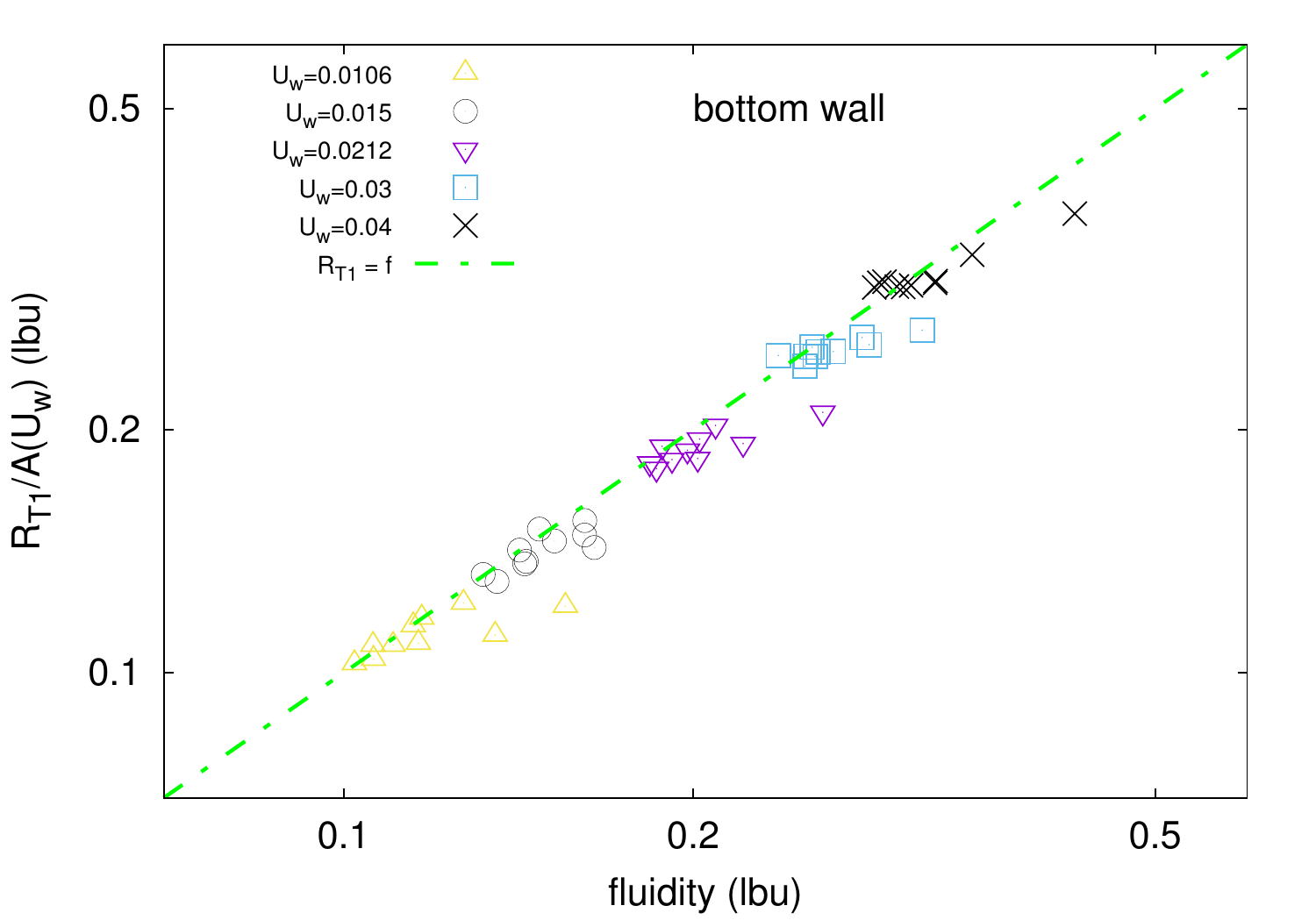}
\caption{(Colour online). Data of Figure \ref{scatter} are here rescaled, by assuming a $U_{\mbox{\tiny{w}}}$-dependent prefactor $\mathcal{A}$ in (\ref{eq:fluidity-2}). Top panel: by assuming the linear scaling law,  $R_{T1} = \mathcal{A} f$ (dashed-dotted line), we are able to extract the $U_{\mbox{\tiny{w}}}$ dependency of the prefactor $\mathcal{A}$ from data in the top slab: a log-log plot of the prefactor $\mathcal{A}$ against the top wall velocity $U_{\mbox{\tiny{w}}}$ is displayed in the inset. The dotted line corresponds to the equation $\mathcal{A}(U_{\mbox{\tiny{w}}}) = 0.017\,U_{\mbox{\tiny{w}}}^{-4/5}$. Bottom panel: data in the bottom slab are rescaled with the prefactor $\mathcal{A}(U_{\mbox{\tiny{w}}})$ computed for the data in the top slab.\label{scatter-rescaled}}
\end{center}
\end{figure}

To make progress, we now report on the role of the roughness on the fluidisation properties of the material in steady Couette flows (i.e. keeping $U_{\mbox{\tiny{w}}}$ constant). We impose that the continuous phase wets perfectly the walls, $\theta_{\mbox{\tiny{c}}}=0^{\circ}$, whereas for the dispersed one $\theta_{\mbox{\tiny{d}}}=180^{\circ}$, thus inducing an effective slip \cite{Bonn2015} on the smooth wall, which is suppressed by the geometrical roughness, as also evident in the rheological curve in the bottom panel of Figure \ref{fig:Sigma_vs_Pxy}. This observation is confirmed by the velocity profiles, shown in the bottom panel of Figure \ref{velprofs-L256-W120} for $(g/d, w/d) = (4.0, 3.5)$.  A small roughness ($h/d=0.4$) reduces the slip; however, as $h$ increases up to a value comparable with the droplet size $h/d=1.1$ (red curve), the material is more fluidised, because of favorable sliding of mobile droplets over those {\it caged} inside inter-posts wells, and the velocity at the edge of the roughness increases. For even higher heights (namely $h/d=1.7$), the extra caging confers more rigidity to the system and the velocity decreases again. Notice that the velocity decrease is seen also for thinner posts ($(g/d, w/d) = (6.7, 0.6)$, top panel), although less strikingly. In connection with roughness-induced fluidisation of the material, deviations from the linear profile arise close to the wall, as highlighted in the inset of the bottom panel of Figure \ref{velprofs-L256-W120}. Such behaviour can be interpreted as a signature of non-locality associated to a cooperative flow \cite{KEP,Bouzid}: the fluidity, i.e. the inverse effective viscosity, varies in space according to Equation (\ref{eq:fluidity}). Such cooperativity shows up as a tiny effect looking at the velocity profiles, so we are now going to investigate directly the fluidity field $f$. In Figure \ref{f-l256-w120} we report the profiles of $f(z)$ (for $g,w,U_{\mbox{\tiny{w}}}$ fixed and different $h$), computed as the ratio of the local shear rate $\dot{\gamma}(z)$ over the local shear stress $\sigma(z)$, both {\it averaged} along the stream-wise direction. Without roughness ($h=0$), the fluidity is the lowest and it is almost constant and equal to its bulk value across the whole channel; due to wall slippage, the material tends to perform a ``plug flow'' \cite{Bonn2015}. A rough wall is then required, as it acts as a source of fluidity; for all $h>0$, in fact, $f(z)$ is higher than in the smooth case, and its profile suggests the emergence of spatial dishomogeneities: it decays from the wall value $f_{\mbox{\tiny{w}}}$ to the bulk one $f_{\mbox{\tiny{b}}}<f_{\mbox{\tiny{w}}}$. The dependence of $f_{\mbox{\tiny{w}}}$ and $f_{\mbox{\tiny{b}}}$ on the posts height $h$ is shown in the inset of Figure \ref{f-l256-w120}. $f_{\mbox{\tiny{w}}}$ grows as $h$ increases, until it reaches a maximum at $h \approx d$, when an optimal {\it caging} occurs, and then decreases again, consistently with the results obtained for the velocity profiles. The bulk fluidity $f_{\mbox{\tiny{b}}}$ is essentially constant up to $h/d \approx 1$ and then very slightly increases. From the micro-mechanical point of view, the rough wall generates fluidity, since it triggers the local yielding of configurations of neighbouring droplets. Therefore, it is tempting to check the validity of the KEP relation (\ref{eq:fluidity-2}) in the numerical data. In order to detect T1 events, we track in time the centers of mass of the droplets and of their nearest neighbours (by using the Voronoi tessellation tools provided by the {\it voro++} library \cite{Voronoi}). Then, we compute the rate $R_{T1}(z)$ within bins of size $\sim d$ (summing over time in the stationary state). In Figure \ref{scatter}, we report the scatter plot of $R_{T1}$ against the fluidity $f$ for various sets of simulations (with fixed $g$ and $w$ and variable $h$ and $U_{\mbox{\tiny{w}}}$): to assess whether different physics emerges with different boundary conditions, we probed the system close to the bottom (rough) and top (smooth) walls separately and, correspondingly, we show data restricted to two slabs of width $\sim 3\, d$ each. Both plots give hints of a power law behaviour, but some cautious words are in order. In the top slab, due to the wall slip, the material tends to perform a kind of ``plug flow'', with the fluidity that gets closer to its bulk value across the whole sub-domain; correspondingly, rather than along the linear scaling, the data tend to accumulate in a polydromic way. This effect is less pronounced in the bottom slab. Overall, it is not possible to reconcile all the data with the same scaling law, exhibiting the same prefactor and the same exponent in the fluidity: data ``clusterize'' in different sets, each labelled by a different wall velocity. For each data set (i.e. fixed $U_{\mbox{\tiny{w}}}$), the spread in fluidity is not enough to well discern whether the scaling exponent is 1 or less. Just for comparison, we have also reported the $R_{T1} \propto f^{0.4}$ relation observed in a molecular dynamics study of Poiseuille flows of jammed soft disks \cite{Mansard2013}. The whole data-set can actually be made compliant with a unique master curve only by assuming the $U_{\mbox{\tiny{w}}}$ dependency of the prefactor $\mathcal{A}=\mathcal{A}(U_{\mbox{\tiny{w}}})$ in (\ref{eq:fluidity-2}). In favor of this argument, it is worth recalling that in the limit of slow flow, in which the KEP result is derived, the elastic modulus $G_0$ can be considered constant, but it decreases with the applied shear, if this is large enough \cite{MasonPRL1995}. The $U_{\mbox{\tiny{w}}}$ dependency of the prefactor $\mathcal{A}$ can be computed, for example, for the data in the top slab, upon the assumption of a linear scaling law. This is shown in the inset of the top panel of Figure \ref{scatter-rescaled}. We then use such $\mathcal{A}(U_{\mbox{\tiny{w}}})$ to rescale data in the bottom slab: if $\mathcal{A}(U_{\mbox{\tiny{w}}})$ is a material property, then we should observe a linear scaling law for the data in the bottom slab, without any extra fitting parameter. Upon such operation (see bottom panel of Figure \ref{scatter-rescaled}), all points tend to align along the $R_{T1} \propto  f$ curve, although with slight deviations. In particular, the best scaling is observed for the lowest values of the wall velocity, i.e. when the material is not far from the yield point. As $U_{\mbox{\tiny{w}}}$ increases, deviations arise: the presence of roughness induces a translational asymmetry that may indeed cause a breakdown of the assumptions which the KEP results are built upon.


\begin{figure*}
\hspace{-0.8cm}
\begin{minipage}{1.2\textwidth}

\includegraphics[scale=0.68]{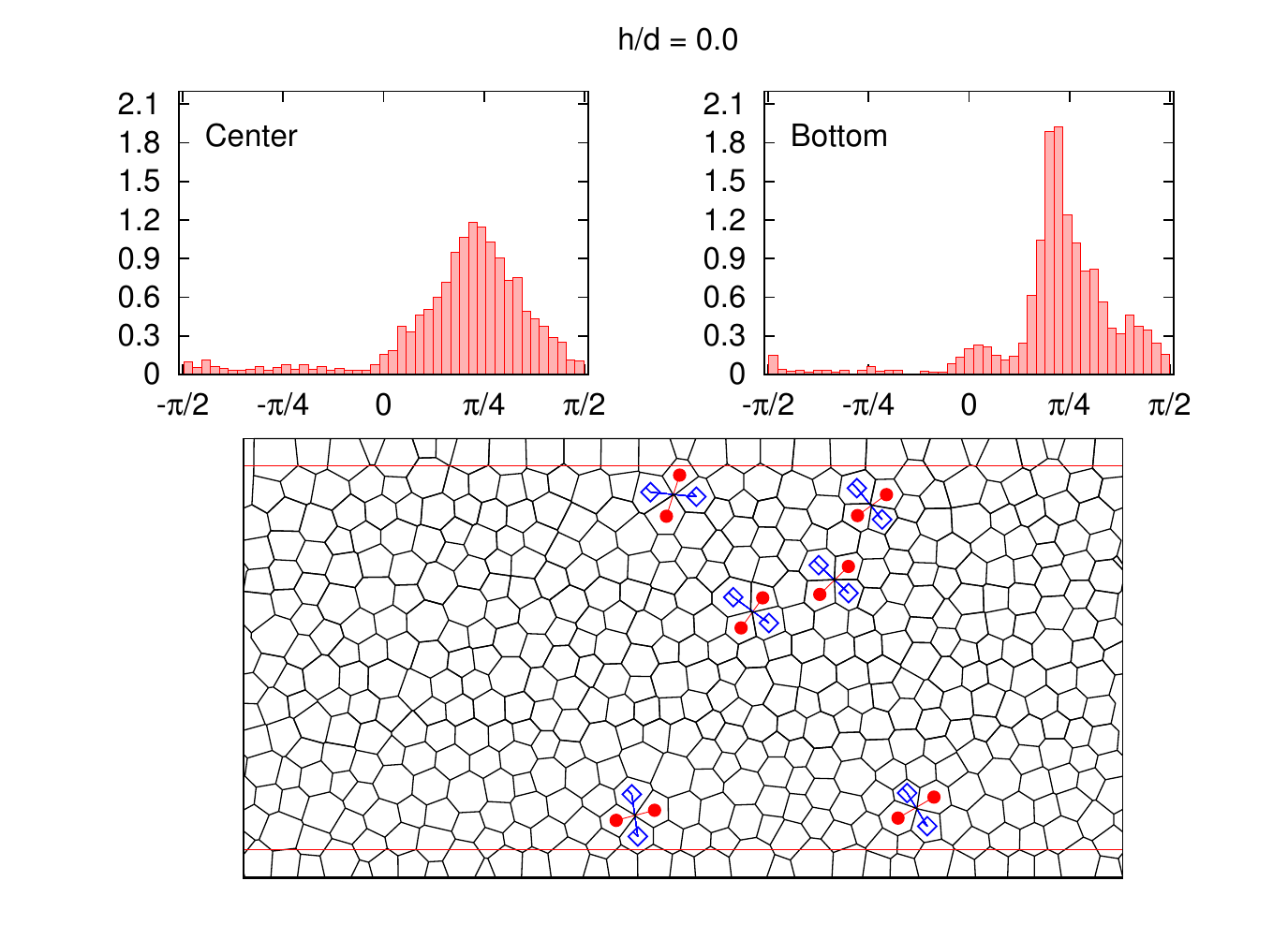}
\includegraphics[scale=0.68]{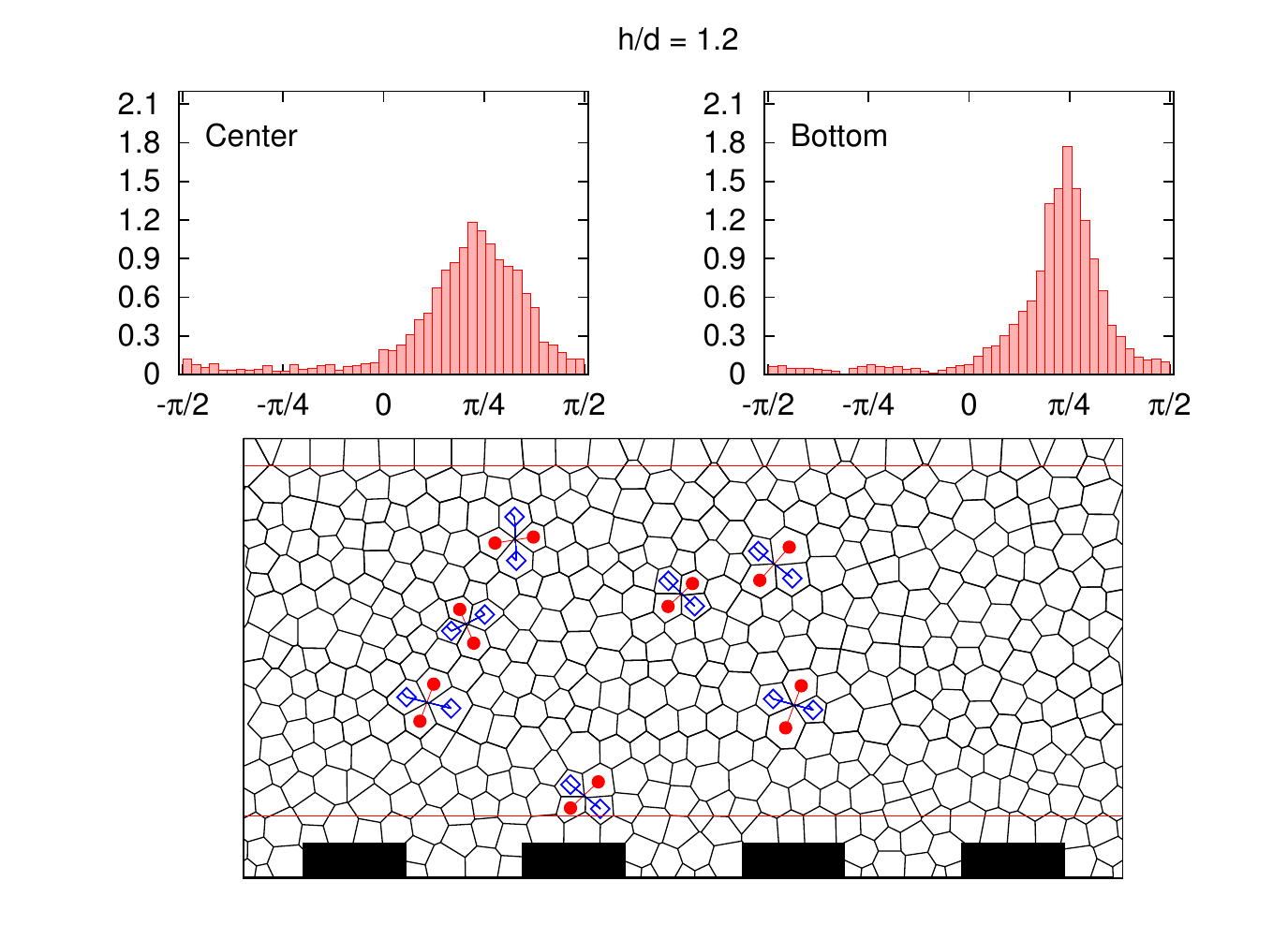}

\end{minipage}
\begin{minipage}{18cm}
\caption{Orientational statistics in a steady Couette cell with smooth ($h/d=0$, left panel) and rough ($h/d= 1.2$, right panel) bottom wall. Following the Voronoi tessellation in time, we are able to identify T1 events and associated ``disappearing'' and ``appearing'' links (see also Figure \ref{sketch}). To highlight the effects of roughness, we report the histograms for the angle of the disappearing links (see text for details and definitions) both in the center of the channel and close to the bottom wall. The horizontal lines show the bounds for the region where the statistical analysis has been carried out. \label{fig:disappearing}}
\end{minipage}

\end{figure*}


So far, the importance of plastic rearrangements has been stressed in terms of their spatial distribution. However, T1 events possess a non-trivial angular structure with a quadrupolar topology \cite{Picard04,Schall07}, hence it is reasonable to characterize also their orientational properties \cite{ourJFM}. Specifically, focusing on the four Voronoi cells involved in a T1, we define as a ``disappearing'' link the segment connecting the centers of the two Voronoi cells that were in contact before the event (and which are then far apart), and as an ``appearing'' link the connector between the other two cells (see also Figure \ref{sketch}). For each event, we then measure the angle between such links and the stream-flow direction. In Figure \ref{fig:disappearing} we report results for the orientational properties of disappearing links in presence of a smooth ($h/d=0$) and rough ($h/d= 1.2$) bottom wall. In the top panel we report histograms for the angle of the disappearing links for 2 distinct regions (each of width $\sim 3\, d$) chosen in the center of the channel and close to the bottom wall. In the bulk of the system, the orientational properties are pretty much the same, with the angle of disappearing links peaked around $\pi/4$. From a qualitative viewpoint this can be understood, since a simple shear flow can be decomposed into an antisymmetric part which provides a rigid-like rotation, and a symmetric part corresponding to an elongational flow along $\pi/4$ \cite{Rioual}. The latter obviously promotes stretching of droplets - and eventually disappearance of links - in such direction. The width of both distributions looks the same, and it is possibly related to the intensity of mechanical noise in the system.  The situation close to the bottom wall, instead, reveals some differences. In both cases (rough and smooth wall), the width of the angular distribution is reduced, which may be taken as an indication that the mechanical noise is less effective in such regions, due to the presence of the confining wall. For the smooth channel, two tiny peaks appear approximately at angles $0$ and $\pi$. We believe this is due to the topological constraint induced by the impenetrable wall, which is absent in the bulk of the system. The wall may indeed favor the formation of ``layers'' of droplets which move aligned in the stream-flow direction. Droplets can leave a layer, thus causing disappearance of a link (with a droplet in an adiacent layer) approximately at a $\pi/2$ angle. Upon entering a new layer, this causes the disappearance of a pre-existing link at an angle equal to $0$. We notice that these two tiny peaks are actually not observed close to the rough wall, which may be taken as a further indication of the extra fluidization induced by the roughness.\\
\section{Conclusions}
Based on mesoscopic numerical simulations, we have explored the properties of boundary conditions for the flow of a jammed soft-glassy material in a rough micro-channel. We considered surfaces with controlled roughness and we have monitored both the averaged velocity profiles and the associated plastic activity. Our data support the picture that both wall friction and surface fluidization are impacted in a non trivial way by the wall roughness. Quantitative insights are also provided in the characterization of the orientational statistics of plastic events, which surely call for a more systematic investigation (e.g., dependence on stress, roughness structure, their correlations) in the coming future. \\
The research leading to these results has received funding from the European Research Council under the European Community's Seventh Framework Programme (FP7/2007-2013)/ERC Grant Agreement no. [279004]

%



\end{document}